\documentclass[twocolumn]{aastex61}
\usepackage{multirow}

\submitjournal{ApJ}

\shorttitle{radio/gamma connection at low radio frequency}
\shortauthors{Fan \& Wu}

\begin{document}

\title{The Radio/Gamma Connection of Blazars from High to Low Radio Frequencies}

\email{fanxl@hust.edu.cn, qwwu@hust.edu.cn}

\author[0000-0002-0786-7307]{Xu-Liang Fan}
\affil{School of Physics, Huazhong University of Science and Technology, Wuhan 430074, China}
\affil{Guizhou Provincial Key Laboratory of Radio Astronomy and Data Processing, Guiyang 550025, China}

\author{Qingwen Wu}
\affiliation{School of Physics, Huazhong University of Science and Technology, Wuhan 430074, China}



\begin{abstract}
  We construct a large sample of $\gamma$-ray blazars with low-frequency radio data using the recent released TGSS AD1 catalog at 150 MHz. The radio/$\gamma$ connections of blazars are compared from 143 GHz to 150 MHz. The radio flux density at all radio frequencies shows strong correlation with $\gamma$-ray flux for blazars, as well as for the two subclasses, FSRQs and BL Lacs. But the correlations get worse from high to low radio frequencies, which indicates that the low-frequency radio emission is the mixture of extended and core components for blazars. In addition, we find that the correlation between 150 MHz radio flux density and $\gamma$-ray flux is more significant for BL Lacs than that for FSRQs. The slope for the luminosity correlation between radio and $\gamma$-ray also get flatter than unity at 150 MHz. These results indicate that the core dominance at 150 MHz for BL Lacs is larger than that for FSRQs. We also compare the radio luminosity from direct TGSS observation and the extended radiation at 150 MHz for blazars. The results show that the ratio between core and extended component at 150 MHz is about 1:1 on average.
\end{abstract}




\section{Introduction} \label{sec:intro}
Blazars are active galactic nuclei (AGNs) whose radiation is dominated by the non-thermal emission from aligned relativistic jets. The spectral energy distribution (SED) of blazars is characterized by two bumps, where the low energy bump is widely accepted to be produced by synchrotron radiation of relativistic electrons. The radiation mechanism of the high energy bump is under debate. The possible mechanisms include the inverse Compton (IC) emission of relativistic electrons (leptonic model), and the hadronic model with protons and pions~\citep{2013ApJ...768...54B}. The IC process, including synchrotron-self Compton (SSC) and external Compton (EC) processes, can usually be successful to reproduce the high energy SEDs of blazars~\citep{2009MNRAS.397..985G, 2010ApJ...716...30A}. Based on the peak frequency of the low energy bump, blazars are usually divided into three subclasses, low synchrotron peaked blazars (LSPs), intermediate synchrotron peaked blazars (ISPs) and high synchrotron peaked blazars (HSPs;~\citealt{2010ApJ...716...30A, 2016ApJS..226...20F}). In general, the peak frequency of both low and high energy bumps are corresponding to each other. That is the LSPs have lower high energy peak, while the high energy spectra is hard for HSPs~\citep{1998MNRAS.299..433F}. The $\gamma$-ray photon spectral index is also found to be correlative with the peak frequency~\citep{2011ApJ...743..171A, 2015ApJ...810...14A}. According to the line emission is present or not, blazars can be divided into flat-spectrum radio quasars (FSRQs) and BL Lac objects (BL Lacs), respectively. In general, most FSRQs are LSPs, while BL Lacs can span much broader range of peak frequency. There are also some researches suggesting that FSRQs and BL Lacs stay in distinct accretion regimes~\citep{2009MNRAS.396L.105G, 2010MNRAS.402..497G, 2014MNRAS.445...81S}.

The radio emission of blazars at GHz frequency is charactized by a flat powerlaw spectrum with $\alpha < 0.5$ ($S_{\nu} \propto \nu^{-\alpha}$;~\citealt{2007ApJS..171...61H}). The radio/$\gamma$ connection of blazars has been explored since the era of Energetic Gamma Ray Experiment Telescope (EGRET) onboard Compton Gamma-Ray Observatory (CGRO; ~\citealt{2001ApJ...556..738J, 2003ApJ...590...95L}). Tight correlations between radio and $\gamma$-ray flux are found for most works based on the Fermi/LAT observations, where the radio data at various GHz frequencies are applied~\citep{2009ApJ...696L..17K, 2010MNRAS.407..791G, 2011MNRAS.413..852G, 2011A&A...535A..69N, 2011ApJ...726...16L, 2011ApJ...741...30A, 2012RAA....12.1475F}. These correlations are usually explained by that the radio and $\gamma$-ray emission is generated by the same population of relativistic electrons. This is also supported by the weakened trend of the radio/$\gamma$ connection from low to high $\gamma$-ray bands~\citep{2011ApJ...741...30A, 2012RAA....12.1475F, 2017A&A...606A.138L}. In spite of many exploration with GHz band or high-frequency radio data, the radio/$\gamma$ connection at low frequency (hundreds MHz), is less explored. Several attempts showed that the radio/$\gamma$ connection at low frequency was worse than that at high frequencies~\citep{2013ApJS..207....4M, 2013ApJS..208...15M, 2016A&A...588A.141G}.

Due to the Doppler boosted emission from core region, the origin of low-frequency radio emission in blazars is under debate. Similar with the large scale radio galaxies, it is believed to be dominated by the extended jet structure, as the low-frequency radio emission from core should be absorbed by synchrotron-self absorption (SSA). However, the spectra between MHz and GHz show a signature of flat spectra for blazars~\citep{2013ApJS..207....4M, 2013ApJS..208...15M}. The low frequency spectra of blazars around hundreds MHz are also flatter than other radio objects~\citep{2016A&A...588A.141G}, which supports that the low-frequency radio emission of blazars is the mixture of extended and core radiation~\citep{2014ApJS..213....3M, 2016A&A...588A.141G}.


Recent years, several new all sky surveys with much higher sensitivity and angular resolution at low radio frequency (around 150 MHz) have been proposed (see Figure 1 of~\citealt{2017A&A...598A.104S}). Several early catalogs of these surveys at low frequency have been released, such as Low-Frequency Array (LOFAR) Two-meter Sky Survey (LoTSS,~\citealt{2017A&A...598A.104S}), GaLactic and Extragalactic All-sky Murchison widefield array (GLEAM,~\citealt{2017MNRAS.464.1146H}) survey and TIFR GMRT (Giant Metrewave Radio telescope) Sky Survey (TGSS,~\citealt{2017A&A...598A..78I}). These new and deeper surveys provide good opportunities to explore the radio emission properties of blazars at low radio frequency. Meanwhile, they are also useful for clarifying the nature of unidentified $\gamma$-ray sources~\citep{2013ApJS..207....4M, 2013ApJS..208...15M, 2014ApJS..213....3M}.

In this paper, we compare the radio/$\gamma$ connection of blazars from high to low radio frequencies with the new low-frequency radio survey, and discussed the origin of the low-frequency radio emission. Section 2 describes the samples and data used in this paper. Section 3 gives the results of radio/$\gamma$ connection and other correlations for different radio bands. In section 4, we explore whether the low-frequency radio luminosity can be used to estimate jet power for blazars. The main conclusions are summarized in section 5. Throughout this paper, we use a $\Lambda$CDM cosmology with $H_0 = 71$ km s$^{-1}$ Mpc$^{-1}$, $\Omega_{\rm M}$=0.27, $\Omega_{\Lambda}$=0.73.

\section{Sample Properties} \label{sec:intro}
The 0.1 --- 100 GeV $\gamma$-ray energy flux of the third Fermi/LAT AGN catalogs (3LAC;~\citealt{2015ApJ...810...14A}) clean sample is taken from the third Fermi/LAT source catalog (3FGL), which is based on the first 48 months data of the Fermi/LAT~\citep{2015ApJS..218...23A}. The clean sample of 3LAC contains 1420 blazars, where only 760 sources have redshift measurements~\citep{2015ApJ...810...14A}. The radio data are derived from three catalogs. For low-frequency radio data, we use the catalog of the TGSS AD1~\citep{2017A&A...598A..78I}. The TGSS survey has covered 90\% of all sky between -53$^\circ$ and +90$^\circ$ declination. The median sensitivity limit is about 3.5 mJy beam$^{-1}$. The astrometric accuracy is about 2\arcsec~in right ascension and declination~\citep{2017A&A...598A..78I}. The GHz radio data are taken from the Radio Fundamental Catalog (RFC; Petrov and Kovalev 2018, in preparation). The RFC provides
precise positions with milli-arcseconds accuracies, and correlated flux densities at baselines 1000 - 8000 km for more than 14000 compact radio sources based on very long baseline interferometry (VLBI) observations~\footnote{http://astrogeo.org/rfc/}. The 143 GHz high-frequency data are derived from the 5th edition of Roma-BZCAT catalogue~\citep{2009A&A...495..691M, 2015Ap&SS.357...75M}, which is based on the observation of PLANCK (Planck Compact Source Catalogue Public Release 1;~\citealt{2014A&A...571A..28P}).

\begin{figure}
\centering
\includegraphics[angle=0,scale=.28]{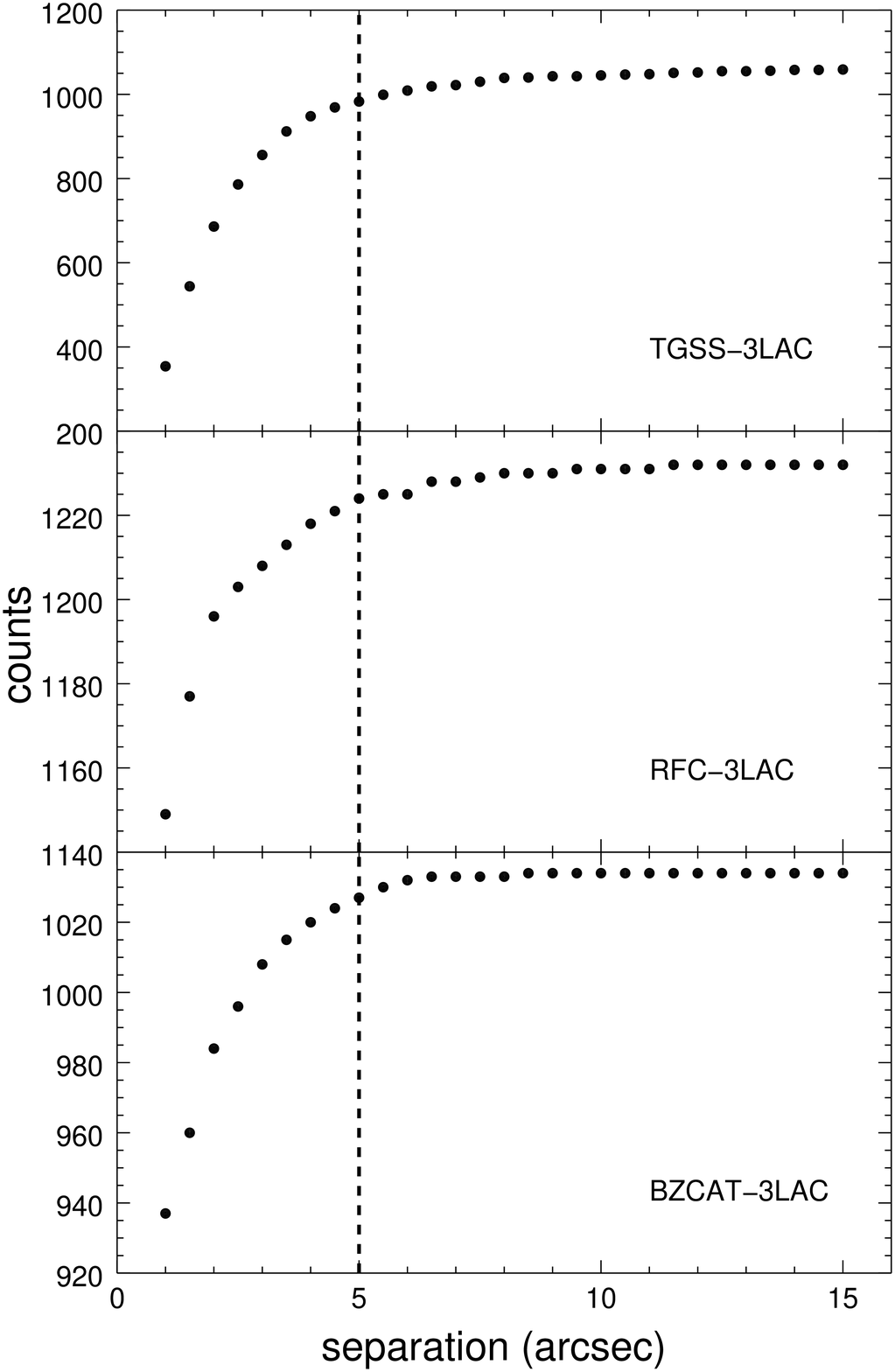}
\caption{The change of source counts with different matching radius. The top, middle, and bottom panels are the results between TGSS AD1, RFC, BZCAT and 3LAC clean sample, respectively. The dashed lines label the matching radius 5~\arcsec. \label{sep}}
\end{figure}

These three radio samples are cross-matched with 3LAC clean sample. The coordinates of the low energy counterparts for $\gamma$-ray sources in 3LAC are used to cross-match with radio catalogs. As the associations of $\gamma$-ray sources are mostly based on the radio catalogs (such as NVSS, SUMSS, CRATES, and CGRaBs,~\citealt{2015ApJ...810...14A}), the expected separations are small. We compare the association results for different matching radius from 1\arcsec~to 15\arcsec, with 0.5\arcsec~step. The results are shown in Figure~\ref{sep}. The counts of associated sources are generally less variable when the matching radius is larger than 5\arcsec. To avoid the spurious associations, we choose 5\arcsec~for all the source associations in our work~\footnote{Note that the selection of the matching radius has little impact on our final samples for statistic analysis, as most sources without radio flux density and redshift data in the cross-matched samples are excluded for statistic analysis.}.

983 $\gamma$-ray sources are found within 5\arcsec~of the TGSS AD1. The detection rate is about 74.0 \% compared with 1328 $\gamma$-ray objects within the TGSS field. As the luminosity is needed in our analyses below, we only consider the sources with redshift measurements. Among the 983 sources, there are 603 sources with redshift measurements, including 364 FSRQs, 197 BL Lacs and 42 AGNs of other types.

Benefited from the VLBI observations of RFC, the total and unresolved (core) emission at GHz can be treated individually. Three frequencies of K band (22 GHz), C band (5 GHz), and S band (2.2 GHz) radio data are applied in this work. There are 127, 315, and 498 sources with redshift measurements after cross-matched with 3LAC clean sample for 22 GHz, 5 GHz, and 2.2 GHz, respectively.

For 143 GHz data, there are 281 blazars with redshift measurements in BZCAT within 5\arcsec~of the associations in 3LAC clean sample, including 222 FSRQs and 46 BL Lacs. The construction of each sample is also summarized in Table~\ref{cor}. For the K-correction when the luminosity is calculated, the spectral index is assumed to be 0.7 for 150 MHz and 0.0 for the high frequencies.
\begin{table*}
\begin{center}
  \caption{The results of correlation analysis. Column 1 gives the frequency for radio data, where core and total components are treated individually for RFC. Columns 2 and 3 are the two parameters which are applied for correlation analysis, respectively. N is the sample number. $\rho$ is the correlation coefficient of the Spearman correlation test. P is the chance probability of no correlation.}
  \setlength{\tabcolsep}{3pt}
  \label{cor}
  \begin{tabular}{cccccccccccc}
  \hline
   &  &  & & All & & & FSRQ & & & BL Lac &  \\
  Radio Frequency & Par A & Par B  & N & $\rho$ & P &  N & $\rho$ & P & N & $\rho$ & P \\
  \hline
   & F$_{143 \rm GHz}$ & F$_{\gamma}$ & 281 & 0.50 & 3.4e-19 & 222 & 0.49 & 8.6e-15 & 46 & 0.59 & 1.7e-05 \\
  143 GHz & L$_{143 \rm GHz}$ & $\Gamma_{\gamma}$ & 281 & 0.22 & 2.2e-04 & 222 & 0.10 & 0.14 & 46 & 0.23 & 0.13 \\
   & F$_{\gamma}$/F$_{143 \rm GHz}$ & $\Gamma_{\gamma}$ & 281 & -0.28 & 1.9e-06 & 222 & -0.25 & 1.9e-04 & 46 & -0.54 & 9.0e-05 \\
  \hline
   & F$_{22 \rm GHz}^{\rm core}$ & F$_{\gamma}$ & 127 & 0.39 & 4.4e-06 & 94 & 0.36 & 3.6e-04 & 24 & 0.19 & 0.38 \\
  22 GHz core & L$_{22 \rm GHz}^{\rm core}$ & $\Gamma_{\gamma}$ & 127 & 0.13 & 0.16 & 94 & -0.006 & 0.95 & 24 & 0.09 & 0.69 \\
   & F$_{\gamma}$/F$_{22 \rm GHz}^{\rm core}$ & $\Gamma_{\gamma}$ & 127 & -0.25 & 0.004 & 94 & -0.20 & 0.06 & 24 & -0.69 & 1.7e-04 \\
  \hline
   & F$_{22 \rm GHz}^{\rm tot}$ & F$_{\gamma}$ & 127 & 0.40 & 2.8e-06 & 94 & 0.41 & 4.7e-05 & 24 & 0.10 & 0.64 \\
  22 GHz total & L$_{22 \rm GHz}^{\rm tot}$ & $\Gamma_{\gamma}$ & 127 & 0.15 & 0.09 & 94 & 0.009 & 0.93 & 24 & 0.13 & 0.56 \\
   & F$_{\gamma}$/F$_{22 \rm GHz}^{\rm tot}$ & $\Gamma_{\gamma}$ & 127 & -0.31 & 3.8e-04 & 94 & -0.29 & 0.005 & 24 & -0.71 & 1.2e-04 \\
  \hline
   & F$_{5 \rm GHz}^{\rm core}$ & F$_{\gamma}$ & 315 & 0.46 & 4.0e-18 & 216 & 0.46 & 1.8e-12 & 83 & 0.64 & 5.3e-11 \\
  5 GHz core & L$_{5 \rm GHz}^{\rm core}$ & $\Gamma_{\gamma}$ & 315 & 0.43 & 1.8e-15 & 216 & 0.15 & 0.03 & 83 & 0.41 & 1.3e-04 \\
   & F$_{\gamma}$/F$_{5 \rm GHz}^{\rm core}$ & $\Gamma_{\gamma}$ & 315 & -0.45 & 6.1e-17 & 216 &-0.29 & 1.8e-05 & 83 & -0.60 & 2.8e-09 \\
  \hline
   & F$_{5 \rm GHz}^{\rm tot}$ & F$_{\gamma}$ & 315 & 0.52 & 4.2e-23 & 216 & 0.52 & 1.1e-16 & 83 & 0.64 & 6.9e-11 \\
  5 GHz total & L$_{5 \rm GHz}^{\rm tot}$ & $\Gamma_{\gamma}$ & 315 & 0.42 & 9.4e-15 & 216 & 0.12 & 0.08 & 83 & 0.41 & 1.2e-04 \\
   & F$_{\gamma}$/F$_{5 \rm GHz}^{\rm tot}$ & $\Gamma_{\gamma}$ & 315 & -0.45 & 2.2e-17 & 216 & -0.29 & 2.0e-05 & 83 & -0.59 & 3.4e-09 \\
  \hline
   & F$_{2.2 \rm GHz}^{\rm core}$ & F$_{\gamma}$ & 498 & 0.41 & 5.4e-22 & 329 & 0.40 & 5.0e-14 & 132 & 0.55 & 7.4e-12 \\
  2.2 GHz core & L$_{2.2 \rm GHz}^{\rm core}$ & $\Gamma_{\gamma}$ & 498 & 0.46 & 4.8e-27 & 329 & 0.10 & 0.08 & 132 & 0.51 & 4.6e-10 \\
   & F$_{\gamma}$/F$_{2.2 \rm GHz}^{\rm core}$ & $\Gamma_{\gamma}$ & 498 & -0.45 & 2.1e-26 & 329 & -0.30 & 2.3e-08 & 132 & -0.67 & 2.9e-18 \\
  \hline
   & F$_{2.2 \rm GHz}^{\rm tot}$ & F$_{\gamma}$ & 498 & 0.43 & 2.2  e-24 & 329 & 0.42 & 8.2e-16 & 132 & 0.54 & 1.8e-11 \\
  2.2 GHz total & L$_{2.2 \rm GHz}^{\rm tot}$ & $\Gamma_{\gamma}$ & 498 & 0.46 & 3.5e-27 & 329 & 0.09 & 0.11 & 132 & 0.51 & 4.3e-10 \\
   & F$_{\gamma}$/F$_{2.2 \rm GHz}^{\rm tot}$ & $\Gamma_{\gamma}$ & 498 & -0.47 & 7.3e-29 & 329 & -0.32 & 2.1e-09 & 132 & -0.65 & 3.9e-17 \\
  \hline
   & F$_{150 \rm MHz}$ & F$_{\gamma}$ & 603 & 0.35 & 6.4e-19 & 364 & 0.29 & 2.6e-08 & 197 & 0.50 & 1.2e-13  \\
  150 MHz & L$_{150 \rm MHz}$ & $\Gamma_{\gamma}$ & 603 & 0.52 & 7.9e-43 & 364 & 0.06 & 0.27 & 197 & 0.42 & 9.8e-10 \\
   & F$_{\gamma}$/F$_{150 \rm MHz}$ & $\Gamma_{\gamma}$ & 603 & -0.29 & 5.4e-13 & 364 & -0.11 & 0.03 & 197 & -0.27 & 1.6e-04 \\
  \hline
  \end{tabular}
\end{center}
\end{table*}

\section{Results And Discussions}
\begin{figure*}
\centering
\includegraphics[angle=0,scale=.3]{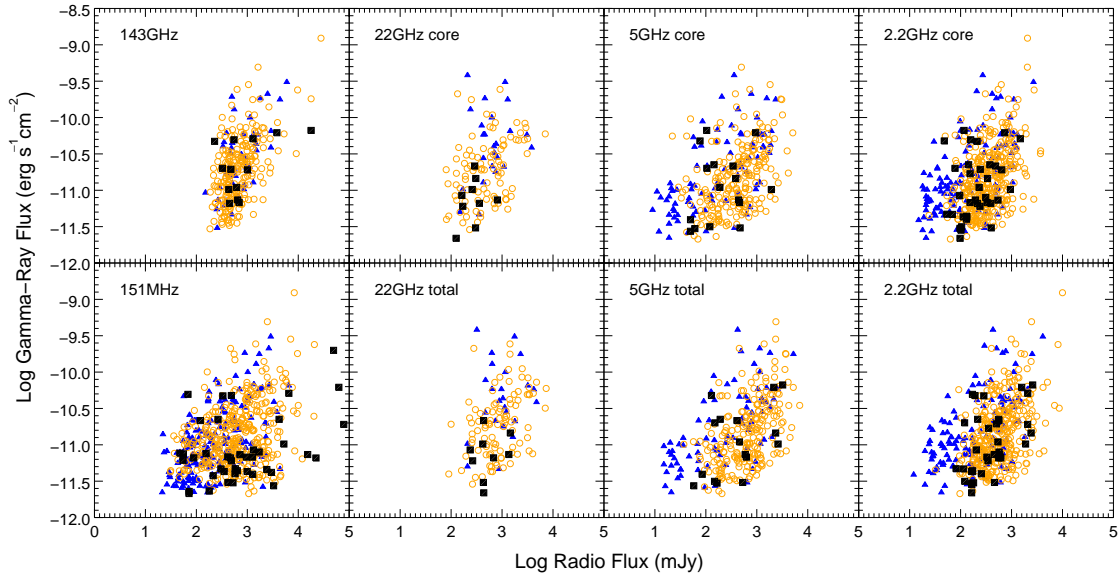}
\caption{The radio/$\gamma$ connection for different radio frequencies. The orange open circles represent FSRQs, blue triangles are BL Lacs, black squares are blazars of uncertain types. \label{rg}}
\end{figure*}

Figure~\ref{rg} shows the flux correlation between radio and $\gamma$-ray fluxes for different radio frequencies. The results of Spearman correlation test for each radio band are given in Table~\ref{cor}. The core and total flux densities of VLBI observations from RFC are treated individually. We also examine the correlation for FSRQs and BL Lacs individually. The 150 MHz radio flux density shows significant correlation with $\gamma$-ray flux, although the correlation coefficient is smaller than higher frequencies. The 143 GHz and 5 GHz radio flux densities show the strongest correlation with $\gamma$-ray flux with the correlation coefficient of 0.50 and 0.52. In addition, the correlation of BL Lacs is generally better than that of FSRQs, except the 22 GHz radio flux density. No correlation is found between 22 GHz radio flux density and $\gamma$-ray flux for BL Lacs. This may be caused by the small sample (24 objects). In particular, the flux correlation at 150 MHz for BL Lacs is much better ($\rho = 0.50$) than that for FSRQs ($\rho = 0.29$). As the radio/$\gamma$ connection is originated from the related physical processes of radio and $\gamma$-ray emission (e.g.,~\citealt{2011ApJ...741...30A}), the more significant correlation for BL Lacs indicates that the core component is more dominant in BL Lacs.

The correlations for core and total emission from VLBI observations show no obvious difference. Fractional reason is that the total VLBI flux density is dominated by core component. The tight connection between the flux density of radio core and $\gamma$-ray flux indicates that the $\gamma$-ray emission is related to the VLBI cores.

\begin{figure*}
\centering
\includegraphics[angle=0,scale=.3]{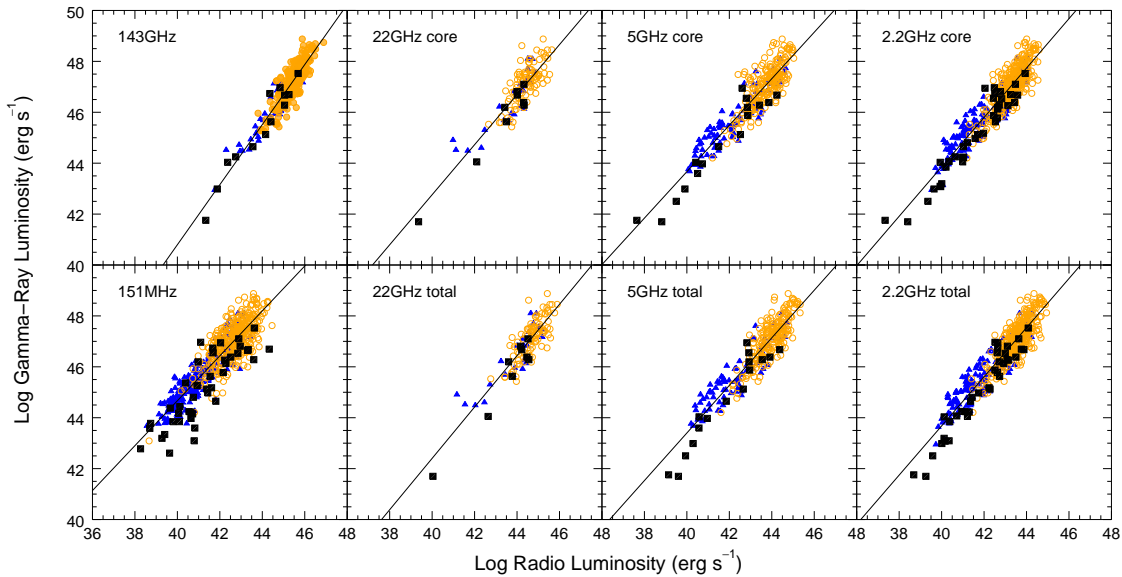}
\caption{The linear correlation between radio and $\gamma$-ray luminosity for different radio frequencies.  The symbols are same with Figure~\ref{rg}. The solid lines show the best fits. \label{lrg}}
\end{figure*}

To clarify the origin for the weaker correlation at low radio frequencies further, we compare the linear relation between radio and $\gamma$-ray luminosity for different radio bands (Figure~\ref{lrg}) with $L_{\gamma} = A L_{\rm radio} + B + \sigma$ (where $\sigma$ is the intrinsic scatter for the linear relation). The Bayesian approach for linear regression proposed by~\citet{2007ApJ...657..116K} are applied to find the linear relations between radio and $\gamma$-ray luminosity. The results of linear fit are summarized in Table~\ref{linear}. The linear relations get flatter from high to low radio frequencies with similar intrinsic scatters. The slopes for GHz (from several to tens) radio luminosity are well consistent with unity~\citep{2011ApJ...742...27L}, which suggest the common origin from the same group of electrons. The slope for low-frequency radio luminosity is about 0.88, which suggests a different origin of low-frequency radio emission.

Considering the radio luminosity at GHz and MHz band, if the spectral index is constant for all sources, the slope would keep similar between different bands. The flatter trend indicates that radio luminosity difference at low luminosity end is larger than that at high luminosity part, i.e, the spectral index between high and low frequency of low luminosity sources is flatter than that of high luminosity sources. This is supported by the direct analyses of blazar spectra at low frequency. The spectral index of FSRQs is slightly larger than that of BL Lacs~\citep{2016A&A...588A.141G}. The flatter spectra of BL Lacs indicates that the core emission is more dominant, which is also supported by the results of flux correlation. The more dominant extended emission, combined with the higher total luminosity of FSRQs at 150 MHz, indicate that the luminosity of extended structures of FSRQs is also higher than that of BL Lacs. This is consistent with their unified version --- radio galaxies, where FR IIs are more luminous than FR Is~\citep{1974MNRAS.167P..31F}. For the steepening trend between GHz and 143 GHz band, a possible explanation is that the variability at high frequency is more intense and the variability amplitude of high luminosity sources, i.e., LSPs is larger than that of low luminosity ones~\citep{2014MNRAS.438.3058R, 2016A&A...596A..45F}.

\begin{table}
\begin{center}
  \caption{The results of linear regression analysis.}
  \setlength{\tabcolsep}{3pt}
  \label{linear}
  \begin{tabular}{cccc}
  \hline
  Radio Frequency & A & B & $\sigma$\\
  \hline
  143 GHz & 1.18$\pm$0.03 & -6.39$\pm$1.40 & 0.41  \\
  22 GHz core & 0.98$\pm$0.05 & 3.44$\pm$2.33 & 0.52 \\
  22 GHz total & 1.01$\pm$0.05 & 2.05$\pm$2.30 & 0.51 \\
  5 GHz core & 0.91$\pm$0.02 & 7.15$\pm$0.97 & 0.50 \\
  5 GHz total & 0.93$\pm$0.02 & 6.01$\pm$0.98 & 0.48 \\
  2.2 GHz core & 0.97$\pm$0.02 & 5.03$\pm$0.75 & 0.45 \\
  2.2 GHz total & 0.97$\pm$0.02 & 5.06$\pm$0.75 & 0.46 \\
  150 MHz & 0.88$\pm$0.02 & 9.38$\pm$0.80 & 0.57\\
  \hline
  \end{tabular}
\end{center}
\end{table}

\begin{figure*}
\centering
\includegraphics[angle=0,scale=.3]{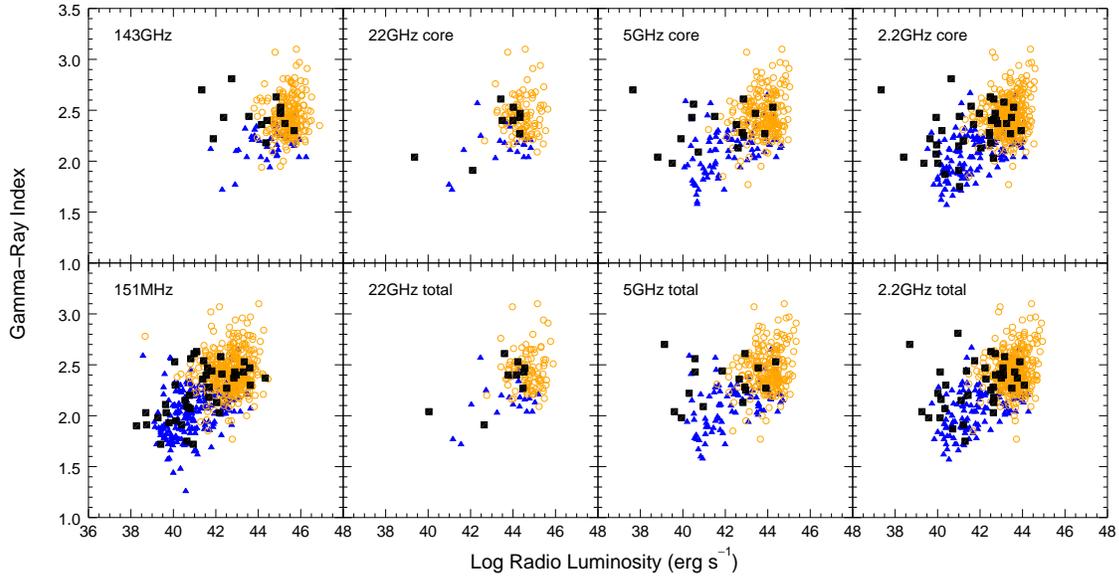}
\caption{The correlation between radio luminosity and $\gamma$-ray index for different radio frequencies.  The symbols are same with Figure~\ref{rg}.  \label{rindex}}
\end{figure*}

\begin{figure*}
\centering
\includegraphics[angle=0,scale=.3]{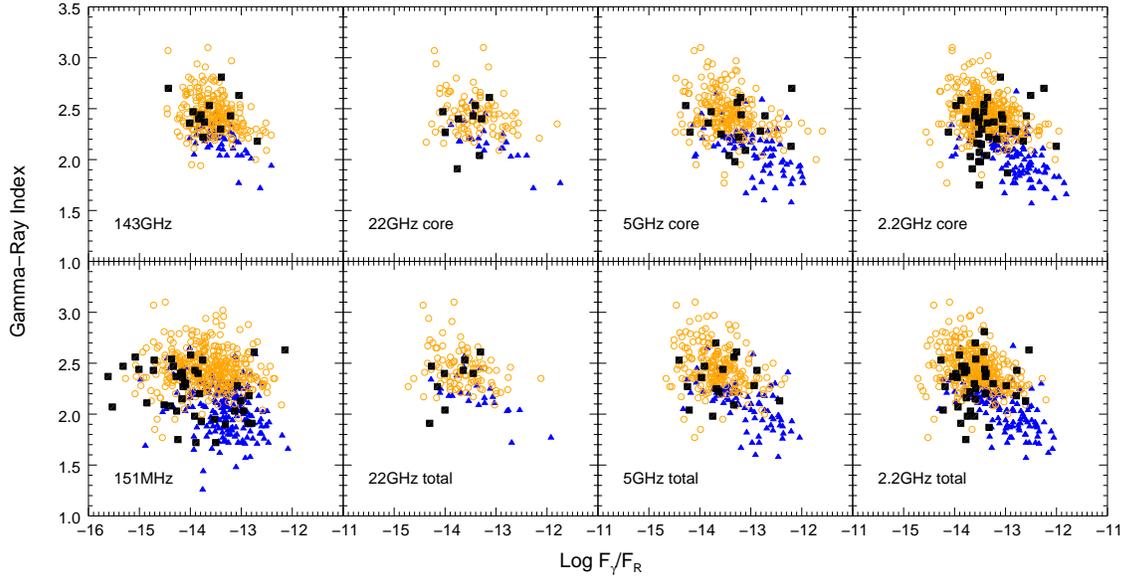}
\caption{The correlation between the flux ratio F$_{\gamma}$/F$_{\rm R}$ and $\gamma$-ray index for different radio frequencies.  The symbols are same with Figure~\ref{rg}. \label{rratio}}
\end{figure*}

We also explore the correlation between the $\gamma$-ray photon spectral index $\Gamma_{\gamma}$ and the radio properties of blazars for different radio frequencies. The results for radio luminosity $L_{\rm radio}$ and flux ratio $F_{\gamma}/F_{\rm radio}$ are plotted in Figure~\ref{rindex} and Figure~\ref{rratio}, respectively. As the $\gamma$-ray spectral index is well correlated with the peak frequency~\citep{2011ApJ...743..171A, 2015ApJ...810...14A}, the correlation between radio luminosity and $\Gamma_{\gamma}$ can be used to explore the so called blazar sequence~\citep{1998MNRAS.299..433F, 2011ApJ...735..108C, 2012MNRAS.420.2899G}. In general, the radio luminosity increase as the $\gamma$-ray spectra get soft. There show a trend that the correlations get better from high to low frequencies (Figure~\ref{rindex}, Table~\ref{cor}). There are two reasons for the correlations. One is the blazar sequence, i.e., the higher-peaked (harder) sources have lower luminosity~\citep{1998MNRAS.299..433F, 1998MNRAS.301..451G}. The other is the position at the SED for a certain observed frequency. For the SEDs with constant shape, the separation between the observed frequency and the peak is larger for HSPs. Thus the higher-peaked (smaller $\Gamma_{\gamma}$) sources would have lower radio luminosity at a certain radio frequency. This effect can also explain why the correlation between radio luminosity and $\Gamma_{\gamma}$ only exists for BL Lacs, as the peak frequency range of FSRQs is much smaller. The correlations disappear at the highest radio frequencies (143 GHz and 22 GHz) even for BL Lacs.

The flux ratio between $\gamma$-ray and radio emission $F_{\gamma}/F_{\rm radio}$, also named $\gamma$-ray dominance or $\gamma$-ray loudness~\citep{2011A&A...535A..69N, 2011ApJ...742...27L}, shows negative correlation with $\gamma$-ray spectral index. The correlation is best for 2.2 GHz and 5 GHz. Similar with the flux correlation, the correlations for BL Lacs is better than those for FSRQs (Figure~\ref{rratio}, Table~\ref{cor}). The similar correlations with 15 GHz and 37 GHz data have been explored by~\citet{2011ApJ...742...27L} and~\citet{2011A&A...535A..69N}, respectively. The $\gamma$-ray loudness is different from the Compton dominance, which is defined by the flux/luminosity ratio between IC bump and synchrotron bump. As mentioned above, the radio flux density at a certain frequency is not only dependent on the blazar sequence, but also on the position at the SED. Similarly, the $\gamma$-ray flux is also dependent on the $\gamma$-ray spectra, which results in the similar range of $\gamma$-ray flux between FSRQs and BL Lacs (Figure~\ref{rg}, also see Figure 8 and 9 of~\citealt{2015ApJ...810...14A}). Thus the negative correlation between $\gamma$-ray loudness and $\Gamma_{\gamma}$, which seems contradict with the blazar sequence (reflected by the negative/positive correlation between Compton dominance and peak frequency/$\Gamma_{\gamma}$), is  comprehensible. As suggested by~\citet{2011ApJ...742...27L}, the significant correlations between the peak frequency/$\Gamma_{\gamma}$ and the $\gamma$-ray loudness indicate that the intrinsic shapes of SEDs are similar for all blazars, including not only the Compton dominance, but also the frequency ratio between two bumps, and width of each bump. Another factor affecting the $\gamma$-ray loudness is the different Doppler transformation for EC and SSC processes~\citep{1995ApJ...446L..63D, 2012ApJ...752L...4M}. However, all these parameters seems correlative with the peak frequency~\citep{1998MNRAS.299..433F, 2013ApJ...763..134F, 2014ApJ...788..179C, 2016RAA....16..173F}. These extra connections bring in large scatters for the correlation with $\gamma$-ray loudness.

\section{Is the 150 MHz radio luminosity a good tracer of the jet power for blazars}
Based on the radio and X-ray observations of radio lobes, several kinetic jet power - 151 MHz radio luminosity relations have been constructed~\citep{1999MNRAS.309.1017W, 2005ApJ...623L...9P, 2010ApJ...720.1066C, 2011ApJ...735...50W, 2013ApJ...767...12G, 2017MNRAS.467.1586I}. These empirical relations are dependent on the assumption that the low-frequency radio emission is dominated by the extended radio structures~\citep{1999MNRAS.309.1017W, 2018MNRAS.475.2768H}. The radio/$\gamma$ connection and the spectral index properties~\citep{2013ApJS..207....4M, 2013ApJS..208...15M, 2016A&A...588A.141G} suggest that the blazar emission at 151 MHz is the combination of core and extended components. Thus can 151 MHz radio luminosity be used to estimate the jet power for blazars?

\begin{figure}
\begin{center}
\includegraphics[angle=0,scale=.35]{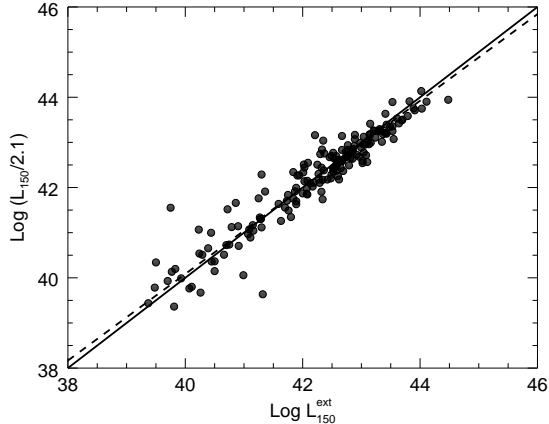}
\caption{The correlation between the extended radio luminosity of 150 MHz extrapolated from 300 MHz and observed 150 MHz radio luminosity corrected by a factor of 2.1. The solid line is the equation line. The dashed line shows the best fit. \label{ext}}
\end{center}
\end{figure}

In order to clarify this, we compare the extended radio luminosity from~\citet{2011ApJ...740...98M} with the 150 MHz data from direct TGSS observations. A spectral index 0.7 is assumed to convert the 300 MHz radio luminosity into 150 MHz. We find that the 150 MHz luminosity from TGSS observation is about 2.1 times higher than that from extrapolation of extended 300 MHz luminosity. This indicates that about half of the radio emission at 150 MHz comes from the extended radio structures, i.e., the ratio between core and lobe is about 1:1. Figure~\ref{ext} shows the comparison between the extended radio luminosity of 150 MHz extrapolated from 300 MHz and observed 150 MHz radio luminosity corrected by a factor of 2.1. A linear regression gives the best fit $\log (L_{150}/2.1) =  0.96 \pm 0.03~(\log L_{150}^{\rm ext} - 40) + 40.09 \pm 0.07$. The slope is well consistent with unity. We also find that this correction makes good unification of jet power distributions between blazars and radio galaxies~\citep{1995PASP..107..803U} for both FSRQs/FR IIs and BL Lacs/FR Is (Fan et al. in preparation).

\citet{2016A&A...588A.141G} decomposed the total flux density as the combination of core and lobe components between 120 MHz and 180 MHz, $S_{\nu} = k_{\rm c} \nu^{-\alpha_{\rm c}} + k_{\rm l} \nu^{-\alpha_{\rm l}} \propto \nu^{-\alpha_{\rm low}}$. By substituting $\alpha_{\rm c}$ = 0.096, $\alpha_{\rm l} = 0.866$ and $\alpha_{\rm low} = 0.57$, they derived $k_{\rm l}/k_{\rm c} = 75$ and the core dominance about 0.63 at 150 MHz. This value is slightly lower than the value here (1.1 on average). The core dominance derived with this method is dependent on the applied spectral index at low radio frequency $\alpha_{\rm low}$ (from 120 MHz to 180 MHz in their case). If the spectral index 0.50 of $\gamma$-ray blazars is applied (see their Table 3), the core dominance at 150 MHz becomes about 0.93, which is generally consistent with our estimations.

It needs to note that the correction with a constant factor is only valid statistically, as the ratio between core and extended component at 150 MHz is variable for single object due to variable Doppler factors, or for various subclasses of blazars (see Section 3). The scatter of low luminosity sources is roughly larger than that of high luminosity sources in Figure~\ref{ext}. This also can be speculated if the core component is more obvious at low luminosity end. And the high luminosity sources is more dominated by extended component at 150 MHz.~\citet{2007MNRAS.381..589M} explored the connection between kinetic jet power and the radio core luminosity of AGNs. They found a significant correlation with a flatter slope $A =  0.54$. After considering the Doppler boosting effect, the slope changed to 0.81. Their results suggest that both the core and extended emission of radio-loud AGNs has tight connection with the kinetic jet power. Thus it is expected that the extended radio luminosity can have a tight relation with the mixture of core and extended radio luminosity, except a scatter from a constant correction factor (2.1 in our case).

\section{Summary}
The radio/$\gamma$ connection of blazars was explored widely in the literatures. In this work, we compare this connection between five different radio frequencies from 143 GHz to 150 MHz. The results confirm that the radio emission at low frequency (150 MHz) is correlated with high energy emission for blazars, which was found with smaller samples in previous works~\citep{2014ApJS..213....3M, 2016A&A...588A.141G}. We confirm the radio/$\gamma$ connection at 150 MHz is worse when compared with higher frequencies, indicates the combination of extended and core emission for blazars at 150 MHz.

The tighter radio/$\gamma$ connection for BL Lacs, combined with the flatter slope of luminosity correlation at 150 MHz, suggest that the core emission at 150 MHz for BL Lacs is more dominant than that for FSRQs.

We compare the 150 MHz radio luminosity of blazars from direct radio observations and extended radio luminosity extrapolated from 300 MHz, and find the fraction of core to extended radio emission is about 1:1 on average. After a correction of a factor of about 2.1, the 150 MHz radio luminosity can be a good tracer of jet power for blazars statistically.

\acknowledgments
We are grateful to the anonymous referee for useful
comments and suggestions. We thank the kindly permission for the usage of RFC data from the RFC Collaboration. This research is supported by National Natural Science Foundation of China (NSFC; grants 11573009 and 11622324) and Guizhou Provincial Key Laboratory of Radio Astronomy and Data Processing (KF201810).

%

\vspace{5mm}
\facilities{GMRT (TGSS AD1), Fermi/LAT (3LAC \& 3FGL)}

\bibliographystyle{aasjournal}
\bibliography{bib}

\end{document}